\documentclass[a4paper]{article}
\usepackage{graphicx}
\usepackage{amsfonts}
\usepackage{amsmath}
\usepackage{amssymb}
\usepackage[english]{babel}
\usepackage[latin1]{inputenc}

\title{BEC and antiparticles in a magnetized neutral vector boson gas at any temperature}

\author{
	LC. Su\'arez-Gonz\'alez$^1$\footnote{lismary@icimaf.cu;lismarydelacaridad@gmail.com}, 
	G. Quintero Angulo$^2$,\\
	A. P\'erez Mart\'{\i}nez$^{1}$ and 
	H. P\'erez Rojas$^1$}

\date{
	$^1$ Instituto de Cibern\'etica Matem\'atica y F\'{\i}sica (ICIMAF),\\
	Calle E esq 15 No. 309 Vedado, La Habana, CP 10400, Cuba\\[5pt]
	$^2$ Facultad de F\'isica, Universidad de La Habana, San L\'azaro y L, CP 10400, Cuba}

\begin{document}

\maketitle

\begin{abstract}
	We study the  thermodynamic properties of a relativistic magnetized neutral vector boson gas at any temperature. We analyse the effect of temperature as well as antiparticles in Bose-Einstein condensation. Astrophysical implications are discussed.
\end{abstract}


\section{Introduction}

One of the most challenging problems of modern physics consists of exploring the behaviour of matter under extreme conditions -- supra-nuclear densities and the presence of strong magnetic fields --, and the determination of the corresponding equations of state (EoS)~\cite{weber2017pulsars}. From the experimental and theoretical point of view, the problem is complex since there are not yet any experiment which pushes matter beyond nuclear saturation density. However, there exist in nature stable and extremely dense stellar configurations that contain matter in one of the denser forms found in the Universe. Astrophysical environments are this way the best scenarios we have to study super-dense matter. In particular, Neutron Stars (NS) are excellent natural laboratories for these studies: they are objects of practically infinite lifetime whose densities can be an order of magnitude higher than those found in atomic nuclei~\cite{weber2017pulsars}, and whose magnetic fields can reach values of up to $(10^9-10^{15})G$ in their surface and $B\sim10^{18}G$ in their core~\cite{lattimer2007neutron}.

When neutron stars are formed they have interior temperatures of the order of $T \sim10^{11}$K (about $10$MeV). For a pure NS, the low temperature limit ($T<<m$) is a valid approximation for description of its structure since the mass of neutrons is $m_n\sim938$ MeV~\cite{schmitt2010dense}. However, at present we know that NS are not only made up of neutrons as originally proposed. In fact, numerous particle processes (ranging from hyperon population and quark deconfinement to the formation of boson condensates) may be competing with each other in the core of a NS~\cite{daicic1993magnetized}. That is why, depending on the type of particle's configuration, it may not be appropriate to take the low temperature approximation to model this kind of star~\cite{daicic1993magnetized}. 

This is the case for instance of positronium that might form in the magnetosphere \cite{wunner1979decay} of the star; with a mass $m\sim1$MeV, the low temperature limit 
previously quoted
can not be taken. But it might also be the case for heavier vector bosons (like neutron-neutron pairs formed in the core of the star\cite{yakovlev2011cooling,chavanis2012bose}) because, due to the occurrence of Bose-Einstein condensation, bosonic gases are more sensitive to temperature than fermionic ones. Since NS are strongly magnetized objects it is also expected that the magnetic properties of the particles that compose the star have an important influence on their phenomenology and structure.

Thermodynamics of relativistic magnetized vector boson gases in the low temperature limit have been studied  in~\cite{ROJAS1996148, Khalilov1997,Khalilov1999,PEREZROJAS2000,angulo2017thermodynamic}.  However, we already saw that the applications of these studies to astrophysical environment might be limited depending on the kind of particle's configurations. Therefore, the purpose of the present work is  to extend the study of the magnetized neutral vector boson gas (NVBG) to all temperature, in order to provide equations of state (EoS) that allows more general and accurate descriptions of astrophysical objects and related phenomena. In this sense, the present paper can be seen as an extension of~\cite{suarez2019non,angulo2017thermodynamic} in which the magnetized neutral vector boson gas was studied for $T<<m$.

Going beyond the low temperature limit not only allows us to obtain more general descriptions of these compact objects, but also to study the corresponding contribution of antiparticles, usually neglected when taking the previously mentioned temperature limit. At high enough temperature the system can have energy for pair production.
The influence of antiparticles in relativistic boson gases is not only important in astrophysics, in fact, it can be of great interest in the final stage of heavy ion collisions because a large number of high-energy hadrons that are produced in these processes could be approximately described as a boson gas~\cite{su2008thermodynamic}.

The paper is organised as follows. In section 2 the particle density for a magnetized neutral vector boson gas  is calculated at any temperature. Section 3 is dedicated to study the contribution of the antiparticles at zero magnetic field. In section 4 the effect of antiparticles and the magnetic field in the Bose-Einstein condensate is studied and compared with the non-relativistic case and the low temperature limit. Finally, concluding remarks are given in section 5. Calculations were done for a spin-1 boson of mass, $m = 2m_n$, where $m_n$ is the neutron mass.

\section{Magnetized vector boson gas at any temperature}\label{sec2}
In this section we'll  compute the particle density for a magnetized neutral vector boson gas at any temperature. For this we start from
the thermodynamic potential per unit volume of a relativistic vector boson gas interacting with a constant and uniform magnetic field $\vec{B}=(0,0,B)$ that was obtained in~\cite{angulo2017thermodynamic}
\begin{equation}\label{eq1}
\Omega^{\pm}=\Omega_{st}+\Omega_{vac} \, ;
\end{equation}
in this expression, $\Omega_{st}$ is the statistical contribution of bosons-antibosons and has the form
\begin{eqnarray}\label{eq2}
\Omega_{st}(\mu,T,b) & \! = \! & -  \sum_{s} \sum_{n=1}^{\infty} \! \bigg(\frac{y_0^2T^2}{2 \pi^2}\frac{e^{n\mu/T}+e^{-n\mu/T}}{n^2}K_2(ny_0/T)\\
& + & \frac{\alpha T}{2 \pi^2} \frac{e^{n\mu/T}+e^{-n\mu/T}}{n}\int_{y_0}^{\infty}\frac{x^2}{\sqrt{x^2+\alpha^2}}K_1(nx/T)\bigg) \, , \nonumber 
\end{eqnarray}
where $\mu$ denotes the chemical potential, $T$ represents the temperature, $s=-1,0,1$ characterises the spin states, $b=B/B_c$ with $B_c=m/2k$ (k is the magnetic moment of the bosons), $y_0=m\sqrt{1-sb}$ and $\alpha=mbs/2$. The second term of Eq.~(\ref{eq1}) is the vacuum contribution and reads
\begin{eqnarray}\label{eq3}
\Omega_{vac}(b) & = & -\frac{m^4}{288\pi}(b^2(66-5b^2)-3(6-2b-b^2)(1-b)^2\\
&&  ln(1-b) -3(6+2b-b^2)(1+b^2)ln(1+b)). \nonumber 
\end{eqnarray}

The particle density of an ideal Bose system with pair production is given by
\begin{equation}\label{eq4}
\rho=\rho^{+}-\rho^{-}=-\frac{\partial \Omega^{\pm} }{\partial \mu}.
\end{equation}

In this work, unlike~\cite{angulo2017thermodynamic} in which the limit $T<<m$ where taken in Eqs.~(\ref{eq2}) and (\ref{eq3}) before any calculation, we will derive directly Eq. (\ref{eq2}) with respect to the chemical potential $\mu$ in order to get an expression for the particle density valid for all temperature
\begin{eqnarray}\label{eq5}
\rho^{+} = \sum_{s} \sum_{n=1}^{\infty} \bigg(\frac{y_0^2T}{2 \pi^2} \frac{e^{n\mu/T}}{n}K_2(ny_0/T)
 + \frac{\alpha}{2 \pi^2} \frac{e^{n\mu/T}}{n}\int_{y_0}^{\infty}\frac{x^2}{\sqrt{x^2+\alpha^2}}K_{1}(nx/T)\bigg) \, ; 
\end{eqnarray}
\begin{eqnarray}\label{eq6}
\rho^{-} =  \sum_{s} \sum_{n=1}^{\infty} \bigg(\frac{y_0^2T}{2 \pi^2} \frac{e^{-n\mu/T}}{n}K_2(ny_0/T)
 + \frac{\alpha}{2 \pi^2} \frac{e^{-n\mu/T}}{n}\int_{y_0}^{\infty}\frac{x^2}{\sqrt{x^2+\alpha^2}}K_{1}(nx/T)\bigg).
\end{eqnarray}

One of the most outstanding properties of bosonic systems, is the occurrence of Bose-Einstein condensation(BEC). The mathematical condition for BEC to occur is that the chemical potential equals the minimum level of energy accessible by a particle $(\mu=\epsilon_{min})$. The magnetized bosons spectrum is
\begin{eqnarray}
\epsilon(p_3,p_{\perp},b,s)=\sqrt{m^2+p_3^2+p_{\perp}^2-msb\sqrt{p_{\perp}^2+m^2}} \, , 
\end{eqnarray}
where $p_{\perp}$ is the momentum component perpendicular to the magnetic field and $p_3$ is the momentum component parallel to it. The ground state energy of the neutral spin-1 boson is~\cite{angulo2017thermodynamic}
\begin{equation}
\epsilon(p_3=p_{\perp}= 0,b,s=1)=m \sqrt{1-b} \, . 
\end{equation}
Equation (\ref{eq4}) is valid only when $|\mu|<m \sqrt{1-b}$ and is interpreted as the particle density of the excited states. To have a more general equation, one must also take into account the particles that are in the ground state~\cite{haber1981thermodynamics}. Then, the particle density can be written as $\rho=\rho_{gs}+\rho^{+}-\rho^{-}$ where $\rho_{gs}$ is the particle density in the ground state. In the next sections we will study these magnitudes in detail.

\section{Antiparticle contribution}\label{sec3}

\begin{figure}[h!]
	\centering
	\includegraphics[width=0.48\linewidth]{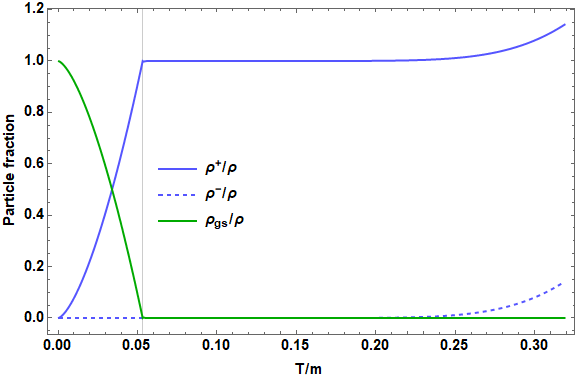}
	\includegraphics[width=0.48\linewidth]{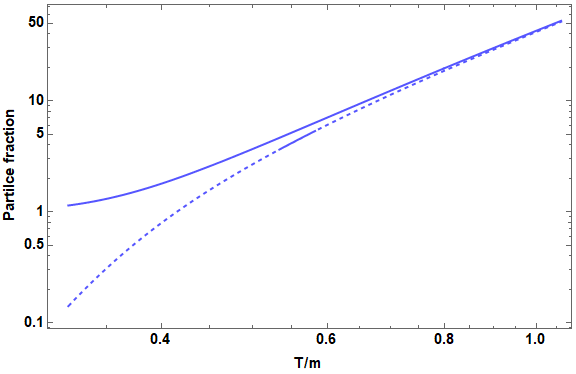}
	\caption{\label{f1} Particle fraction as a function of the temperature. The green line is to the fraction of particles in the ground state. The blue line corresponds to the fraction of particles and the blue line corresponds to the antiparticle fraction. }
\end{figure}

In this section we analyse the antiparticle contribution to the NVBG thermodynamics. Since antiparticle presence is mainly related to temperature, we will study particle-antiparticle density at $B=0$. To compute the thermodynamic potential of a relativistic NVBG at zero magnetic field we start form the boson spectrum $\epsilon(p)=\sqrt{p^2+m^2}$, being $\vec{p}$ the total momentum. For such a gas, the density of states $g(\epsilon)$ is
\begin{eqnarray}\label{eq7}
g(\epsilon)=\sum_{s}\sum_{\vec{p}}\delta \bigl[\epsilon-\sqrt{p^2+m^2}\bigr]
=\frac{4\pi V}{(2\pi)^3}(2s+1)\;2\epsilon\;\sqrt{\epsilon^2-m^2},\;\;\; |\epsilon|<m.
\end{eqnarray}
Using  Eq.~(\ref{eq7}) the thermodynamic potential per unit volume can be written as
\begin{equation}\label{eq8}
\Omega^{\pm}(\mu,T,B)=-T\int_m^\infty d\epsilon g(\epsilon)\;ln \bigl[f_{BE}^{\pm} (\epsilon,\mu)\bigr], \;\;\forall\; |\mu|<\epsilon,
\end{equation}
where 
\begin{equation}
f_{BE}^{\pm} = \bigl[(1-e^{\frac{\mu-\epsilon}{T}})(1-e^{-\frac{\mu+\epsilon}{T})}\bigr]^{-1} \, ,  
\end{equation}
is the the Bose-Einstein distribution function.
Using moreover
\begin{equation}
ln(1-x)=-\sum_{n=1}^{\infty}\frac{x^n}{n} \, ,  
\end{equation}
and 
\begin{equation}
\int_u^{\infty}x(x^2-u^2)^{\nu-1}e^{-\alpha x}dx=2^{\nu-\frac{1}{2}}(\sqrt{\pi})^{-1}u^{\nu+\frac{1}{2}}\Gamma(\nu)K_{\nu+\frac{1}{2}} \, , 
\end{equation}
the thermodynamic potential becomes
\begin{eqnarray}\label{eq10}
\Omega^{\pm}=-3\frac{m^2T^2}{2 \pi^2}\sum_{n=1}^{\infty}\frac{e^{n\mu/T}+e^{-n\mu/T}}{n}K_2(nm/T),
\end{eqnarray}
where $K_{\alpha}$ is the MacDonald function of order $\alpha$. It is easy to verify that the thermodynamic potential Eq.~(\ref{eq2}) is reduced to Eq.~(\ref{eq10}) when we make the magnetic field equal to zero. By deriving Eq.~(\ref{eq10}) with respect to $\mu$, we obtain the following expressions for the particle ($\rho^{+}$) and antiparticle ($\rho^{-}$)  densities
\begin{eqnarray}\label{eq11}
\rho^{+}=3\frac{m^2T}{2 \pi^2}\sum_{n=1}^{\infty}\frac{e^{n\mu/T}}{n}K_2(nm/T),
\end{eqnarray}

\begin{eqnarray}\label{eq12}
\rho^{-}=3\frac{m^2T}{2 \pi^2}\sum_{n=1}^{\infty}\frac{e^{-n\mu/T}}{n}K_2(nm/T).
\end{eqnarray}
Eqs.~(\ref{eq11}) and (\ref{eq12}) are in agreement with the obtained by~\cite{elmfors1995condensation, beckmann1979bose}.
For a fixed value of the total density $\rho$ we can compute the fraction of bosons-antibosons as a function of the temperature and also the particle fraction in the ground state. Figure~\ref{f1} shows the fraction of particles $\rho^+/\rho$, antiparticles $\rho^-/\rho$ and particles in the ground state $\rho_{gs}/\rho$  for a fixed value of the total particle density $\rho=1.67 \times 10^{15}g cm^{-3}$.

The bottom panel of Figure \ref{f1} is the temperature {\it continuation} of the upper panel; this separation was made so that the effects we want to highlight could be better observed. As we can see from the upper panel of Figure \ref{f1}$\rho_{gs}=0,\; \forall T>T_c$, which corresponds to a free gas, while $\rho_{gs} \neq 0$,\;$ \forall T<T_c$ corresponding to a state where free gas and condensate coexist.
On the other hand, we can see from this figure that the contribution of antiparticles begins to be important for $T\simeq 1/4m$ while it becomes of the same order of the particle fraction for $T\geq m$. In this range, the presence of antiparticle degrees of freedom can have an important influence on the thermodynamic properties of the gas.
In the next section we study in more depth the antiparticles effects on Bose-Einstein condensation.

\section{Bose-Einstein condensation}\label{sec4}

To study the effect of the presence of antiparticles and magnetic field on the Bose-Einstein condensation, we compare the critical curve obtained with the 
widespread expression for all temperatures (Eq.~\ref{eq130}) with the corresponding non-relativistic limit studied by \cite{suarez2019non} and with the low temperature limit studied by~\cite{angulo2017thermodynamic}.

\begin{figure}[h!]
	\centering
	\includegraphics[width=0.48\linewidth]{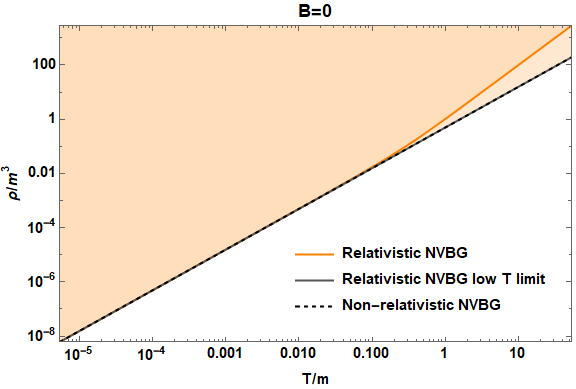}
	\includegraphics[width=0.48\linewidth]{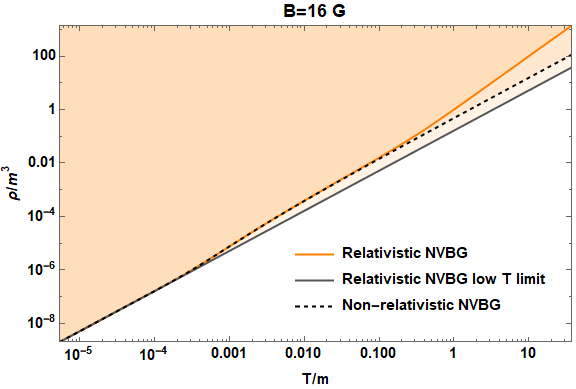}
	\caption{\label{f2} BEC Phase Diagram. The white region corresponds to the state of free gas. The orange shaded region correspond to  the BEC state. The lines correspond to the curve of $\rho_c(b),T_c(b)$ in different approaches to the neutral vector bosons gas description.}
\end{figure}

In a Bose gas, the condensed state can be reached in two ways: by lowering the temperature below the critical temperature, $T_c$, for a fixed value of density, $\rho$, or by increasing the density above the critical density, $\rho_c$, for a fixed value of temperature. A set of values of ($T_c$,$\rho_c$) defined a critical curve which separates the condensed region from the non-condensed one.
This curve depends in addition on the magnetic field ($T_c(b)$,$\rho_c(b)$) and is obtained by evaluating the particle density in the condition for the condensate formation, $\mu = m \sqrt{1-b}$

\begin{eqnarray}\nonumber\label{eq130}
\rho_c(b)&=&\sum_{s}\frac{y_0^2T}{2 \pi^2}\sum_{n=1}^{\infty}\frac{e^{nm\sqrt{1-b}/T}-e^{-nm\sqrt{1-b}/T}}{n}K_2\bigg(\frac{ny_0}{T}\bigg)\\\nonumber
&+&\sum_{s}\frac{\alpha}{2 \pi^2}\sum_{n=1}^{\infty}\frac{e^{nm\sqrt{1-b}/T}-e^{-nm\sqrt{1-b}/T}}{n}\\
&\times&\int_{y_0}^{\infty}\frac{x^2}{\sqrt{x^2+\alpha^2}}K_{1}\bigg(\frac{nx}{T}\bigg) \, . 
\end{eqnarray}
Figure \ref{f2} shows the BEC phase diagram for $B=0$ (top panel) and at $B=10^{16}$ G (bottom panel). The orange line corresponds to the curve of $\rho_c(b)$ given by Eq.~\ref{eq130}. The black line is the critical curve in the low temperature limit\cite{angulo2017thermodynamic}, while the dashed black line corresponds to the non-relativistic NVBG~\cite{suarez2019non}.

At $B=0$  there is no notable difference in the behaviour of the critical curve for the relativistic NVBG at low temperature and  the non-relativistic NVBG.  The critical curve of the relativistic NVBG begins to differentiate from the other two around $T \sim m$ signalling the presence of the antiparticles at those temperatures. 
At $B=10^{16}$ G the critical curve of the relativistic NVBG in the low temperature limit start to be different from the other cases around $T \sim 10^{-3}m$, indicating that this limit is not entirely correct above those temperatures.  The relativistic and non-relativistic NVBG begin to have a different behaviour at $T \sim m$. This difference is given by the presence of antiparticles.
It is perhaps worthwhile to note that the results in Figure~\ref{f2} are in agreement with the condition for a Bose gas to condense at relativistic temperatures, $T_c>>m$, which is $ \rho_c>>m^3$~\cite{haber1982finite}.

\section{Conclusions}\label{sec5}

We have studied the properties of a magnetized neutral vector boson gas at any temperature. Expressions for particle-antiparticle density valid for any magnetic field and temperature were obtained. Starting from them, we study antiparticle contribution as well as Bose-Einstein condensation.

For temperatures in the order of particle masses and higher ($T \geq m$) antiparticle fraction is not negligible as well as their influence on the thermodynamic properties. This is clearly evidenced when we study the BEC phase diagram in which the behaviour of the critical curves at high temperatures differs between the relativistic and the non relativistic cases, precisely due to antiparticle presence.

On the other hand, we can also conclude that the temperature that defines the validity of the low temperature limit depends on the magnetic field. For zero field this limit is valid for $T<<m$ while for $B=10^{16}$ G is only fulfilled for $T<<10^{-3}m$. This implies that even for paired neutrons this limit is no longer valid at $T\sim 10^{10}$ K which is a temperature that can be found in Neutron Stars. This study give us an insight on the importance of extending the equations of state for the NVBG beyond the low temperature limit.\\



\begin{thebibliography}{99}
	
	\bibitem{weber2017pulsars} Weber F 2017 \textit{Pulsars as astrophysical laboratories for nuclear and particle physics} (Routledge)
	
	\bibitem{lattimer2007neutron} Lattimer J M and Prakash M 2007 \textit{Physics reports} \textbf{442} 109-165
	
	\bibitem{schmitt2010dense} Schmitt A 2010 \textit{Dense matter in compact stars: A pedagogical introduction} vol 811 (Springer)
	
	\bibitem{daicic1993magnetized} Daicic J, Frankel N and Kowalenko V 1993 \textit{Physical review letters} \textbf{71 1779}

	\bibitem{wunner1979decay} Wunner G and Herold H 1979 \textit{Astrophysics and Space Science} \textbf{63} 503-509
	
	\bibitem{yakovlev2011cooling} Yakovlev D G, Ho W C, Shternin P S, Heinke C O and Potekhin A Y 2011 \textit{Monthly Notices of the Royal Astronomical Society} \textbf{411} 1977-1988
	
	\bibitem{chavanis2012bose} Chavanis P H and Harko T 2012 \textit{Physical Review D} \textbf{86} 064011
	
	\bibitem{ROJAS1996148} Rojas H P 1996 \textit{Physics Letters B} \textbf{379} 148-152 (\textit{Preprint} hep-th/9510191)
	
		\bibitem{Khalilov1997} Khalilov V, Ho C L and Yang C 1997 \textit{Modern Physics Letters A} \textbf{12} 1973-1981
		
		\bibitem{Khalilov1999} Khalilov V R and Ho C L 1999 \textit{Phys. Rev. D} \textbf{60}(3) 033003
		
		\bibitem{PEREZROJAS2000} Perez Rojas H and Villegas-Lelovski L 2000 \textit{Brazilian Journal of Physics} \textbf{30} 410-418 ISSN 0103-9733
		
		\bibitem{angulo2017thermodynamic} Angulo G Q, Martinez A P and Rojas H P 2017 \textit{Physical Review C} \textbf{96} 045810
		
		\bibitem{suarez2019non} Suarez-Gonzalez L, Angulo G Q, Martinez A P and Rojas H P 2019 \textit{Journal of Physics: Conference Series} vol 1239 (IOP Publishing) p 012004
		
		\bibitem{su2008thermodynamic} Su G, Chen L and Chen J 2008 J\textit{ournal of Physics A: Mathematical and Theoretical} \textbf{41} 285002
		
		\bibitem{haber1981thermodynamics} Haber H E and Weldon H A 1981 \textit{Physical Review Letters} \textbf{46} 1497
		
		\bibitem{elmfors1995condensation} Elmfors P, Liljenberg P, Persson D and Skagerstam B S 1995 \textit{Physics Letters B} \textbf{348} 462-467
		
		\bibitem{beckmann1979bose} Beckmann R, Karsch F and Miller D E 1979 \textit{Physical Review Letters} \textbf{43} 1277
		
		\bibitem{haber1982finite} Haber H E and Weldon H A 1982 \textit{Physical Review D} \textbf{25} 502
		
\end{thebibliography}


\end{document}